# Solar Radio Bursts and Space Weather

Stephen M. White

*Dept. of Astronomy, University of Maryland, College Park, MD 20742 USA*

**Abstract.** Space Weather is the study of the conditions in the solar wind that can affect life on the surface of the Earth, particularly the increasingly technologically sophisticated devices that are part of modern life. Solar radio observations are relevant to such phenomena because they generally originate as events in the solar atmosphere, including flares, coronal mass ejections and shocks, that produce electromagnetic and particle radiations that impact the Earth. Low–frequency solar radio emission arises in the solar atmosphere at the levels where these events occur: we can use frequency as a direct measure of density, and an indirect measure of height, in the atmosphere. The main radio burst types are described and illustrated using data from the Green Bank Solar Radio Burst Spectrometer, and their potential use as diagnostics of Space Weather is discussed.

## 1. Introduction

Low–frequency (below 100 MHz) solar radio bursts were amongst the first radio phenomena studied by radar scientists who turned their radio equipment to the skies after returning from World War II. In this paper we take advantage of exceptionally clean radio spectra of such bursts acquired with the Green Bank Solar Radio Burst Spectrometer to show exampes of these bursts, and to discuss their relationship with space weather phenomena that are of increasing importance in the modern technological world. We will not attempt an exhaustive review of the history of these phenomena, but refer the reader instead to other more comprehensive works: specifically, the books by Kundu (1965) and McLean and Labrum (1985) for observational reviews, Zheleznyakov (1970), Melrose (1980) and Benz (2002) for the theory of radio emission in these bursts, and Pick (2004) and Schwenn (2006) for the relationship with space weather.

The value of solar radio bursts at low frequencies lies in the fact that they originate in the same layers of the solar atmosphere in which geo–effective disturbances probably originate: the layers where energy is released in solar flares, where energetic particles are accelerated, and where coronal mass ejections (CMEs) are launched. We are confident of this because at low frequencies, most emission takes the form of plasma emission. Plasma emission is a resonant process in which electrostatic Langmuir waves at the electron plasma frequency, $\nu_p = 9000\sqrt{n_e}$ where $n_e$ (cm$^{-3}$) is the ambient electron number density, are driven to very high effective brightness temperatures by coherent interaction with a beam or a loss–cone pitch–angle distribution and then convert to propagating transverse electromagnetic waves at $\nu_p$ or its harmonic, $2\nu_p$. The plasma frequency at the canonical active–region density of $10^9$ cm$^{-3}$ is 300 MHz. Radiation at $\nu_p$ is damped by collisional opacity, which is proportional to $\nu_p^4$, and so this emission is typically seen only at low frequencies (i.e., low densities) or when the ambient plasma is very hot.

For what appears on the surface to be an exotic mechanism, plasma emission is remarkably prevalent in the solar atmosphere, appearing in some form in quite a large fraction of flares (Benz et al., 2005; Benz et al., 2007). It is not the only emission mechanism operating at frequencies below 1000 MHz, however. Optically–thick thermal bremsstrahlung emission dominates the quiet Sun in this range, producing brightness temperatures of order $10^6$ K, while electron cyclotron maser emission, operating at the fundamental of the electron gyrofrequency, $\nu_B = 2.8 \times 10^6 B_{\text{gauss}}$, and at the harmonic $2\nu_B$ (Holman et al., 1980; Melrose and Dulk, 1982), provides a better explanation of the properties of some emissions in this range (Benz, 2004). In this paper we will mostly discuss emission below 100 MHz, where plasma emission is believed to dominate most of the observed burst types.

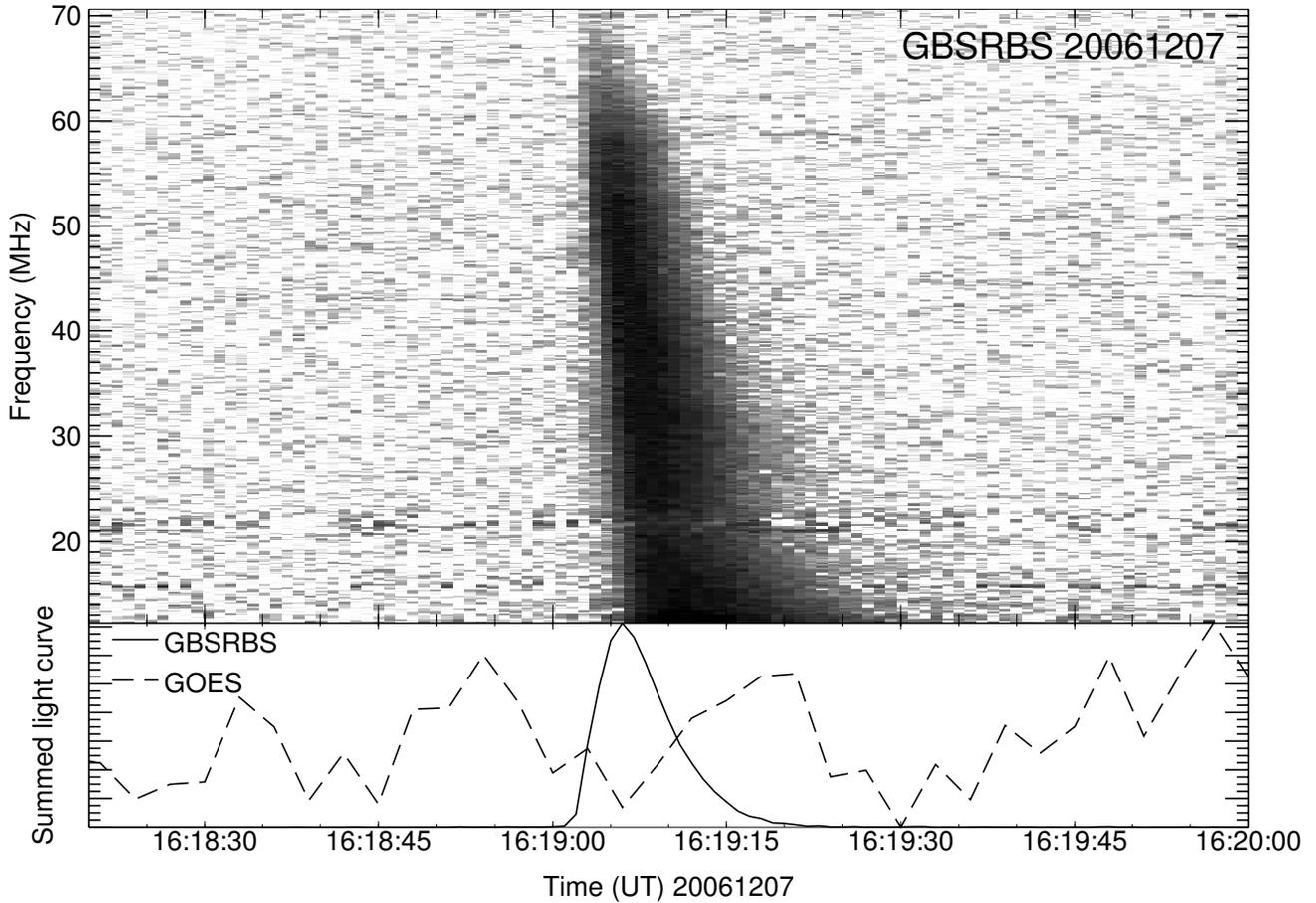

*Figure 1.* A simple isolated Type III burst observed with the low–frequency system of the Green Bank Solar Radio Burst Spectrometer on December 7, 2006. The frequency range on this day was 12 to 70 MHz. This figure uses a format common to the all the plots of dynamic spectra in this paper: in order to exploit the black–and–white display, the display range is inverted so that darker pixels represent brighter emission. The bottom level of the display range is set so that the background noise level is just visible. The light curves in the lower panel of this plot show a frequency–integrated radio light curve (solid line, 1–second time resolution) and for comparison, a soft X–ray light curve from the detectors on the GOES satellites (dashed line, 3–second time resolution). The two light curves are scaled arbitrarily to fit on the same plot: solar flares are generally obvious in the GOES soft X–ray light curves as large increases above the apparent noise level. In this case there was no flare in conjunction with the Type III burst.

## 2. Space Weather

The conditions in the solar wind in the Earth's vicinity are now referred to generically as "Space Weather". These conditions include the solar wind speed and density, magnetic field strength and orientation, and energetic particle levels. They are largely controlled by the Sun, which is the source of the solar wind as well as of coronal mass ejections that impact the Earth with high densities and magnetic field strengths travelling at up to thousands of km s$^{-1}$, and of flares and eruptions that accelerate particles to damagingly high energies and send them towards the Earth. The Earth's magnetosphere and atmosphere have historically protected us from most of the potentially damaging effects of Space Weather. The magnetosphere's closed magnetic field lines cushion us from the shocks provided by changing conditions in the solar wind, and deflect much of the damaging ionized radiation flux from the Sun. The atmosphere absorbs most of the large flux of ionizing ultraviolet, extreme ultraviolet and soft X–ray photons produced by solar flares that would otherwise damage biological cells, and life has adapted to survive the resulting conditions at the surface of the Earth.

But the increasing reliance of modern society on technologically advanced electronic systems has resulted in vulnerability to electromagnetic and particle influences from the sources external to the Earth. The inter–connected international power grids contain long–range electrical connections that are susceptible to the large–scale electric fields generated when the magnetosphere is compressed by disturbances in the solar wind, resulting in voltage and current overloads that can shut down power supplies to millions of people. Long–distance oil and gas pipelines, particularly at high latitudes, are similarly subjected to currents that cause damaging corrosion. Use of global positioning data from satellites is becoming widespread and many industries now depend on such data: air traffic is moving towards relying entirely on such technology. But the technology requires precise timing of radio signals passing through the Earth's ionosphere, where changing conditions caused by Space Weather effects can disrupt measurements. Cell phones rely on communications at microwave frequencies where the Sun can cause harmful interference during flares, disrupting service. Other ground–based radio communication methods require low–absorption paths through the atmosphere that can be destroyed when ionizing radiation from the Sun increases the charged particle densities in the lower ionosphere. With increased astronaut activity expected in coming years as NASA sends missions to the Moon and Mars, there is concern about the possibility of deadly radiation storms occurring during manned missions. There are frequent losses of satellites in low–Earth orbit due to increased drag from the atmosphere during periods of high solar activity when the upper atmosphere is heated by ionizing photon fluxes and expands outwards, while all satellites are susceptible to radiation damage in critical computing components that can result in complete loss of control. The commercial implications of Space Weather are now widely recognized and insurance companies in particular are paying attention to its effects on their industry.

For all of these reasons, the study of Space Weather has become an important practical task in addition to the intellectual value of understanding the physical processes involved. Since most of Space Weather's effects originate in the Sun's atmosphere, any diagnostics there can potentially be valuable. Radio observations sample most of the activity in the Sun's atmosphere and are expected to play an important role in monitoring Space Weather sources.

### 3.  The Green Bank Solar Radio Burst Spectrometer

The Green Bank Solar Radio Burst Spectrometer (SRBS) has been operating routinely since January 2004. Funded by the National Science Foundation under a grant to the National Radio Astronomy Observatory (NRAO: principal investigator, Tim Bastian), SRBS takes advantage of the national radio–quiet zone around the Green Bank Observatory in West Virginia to obtain low–noise spectra of solar radio bursts from the ionospheric cutoff at about 10 MHz up to 1000 MHz. Three frequency ranges are served by different feeds: (a) a long–wavelength dipole antenna, currently the "coat–hanger" type designed by Bill Erickson and tested for the Long Wavelength Array, has been operating from 18–70 MHz since 2004; (b) a 200-1100 MHz feed mounted at the prime focus of a 13.7 m dish (donated by NRAO to the project) has been operating since mid 2005; and (c) a 70–300 MHz feed mounted above the feed support for the 13.7 m dish but pointing outwards was installed in mid 2006 and is currently (early 2007) being commissioned.

Each receiver feeds a commercial spectrum analyzer that sweeps through each frequency range of order 1000 times per second with of order 1000 spectral channels each, and the individual sweeps are currently binned up to 1 second time resolution. The resulting data are very high quality spectra with low background levels. The daily data are processed by background subtraction and are available over the web at `http://gbsrbs.nrao.edu`.

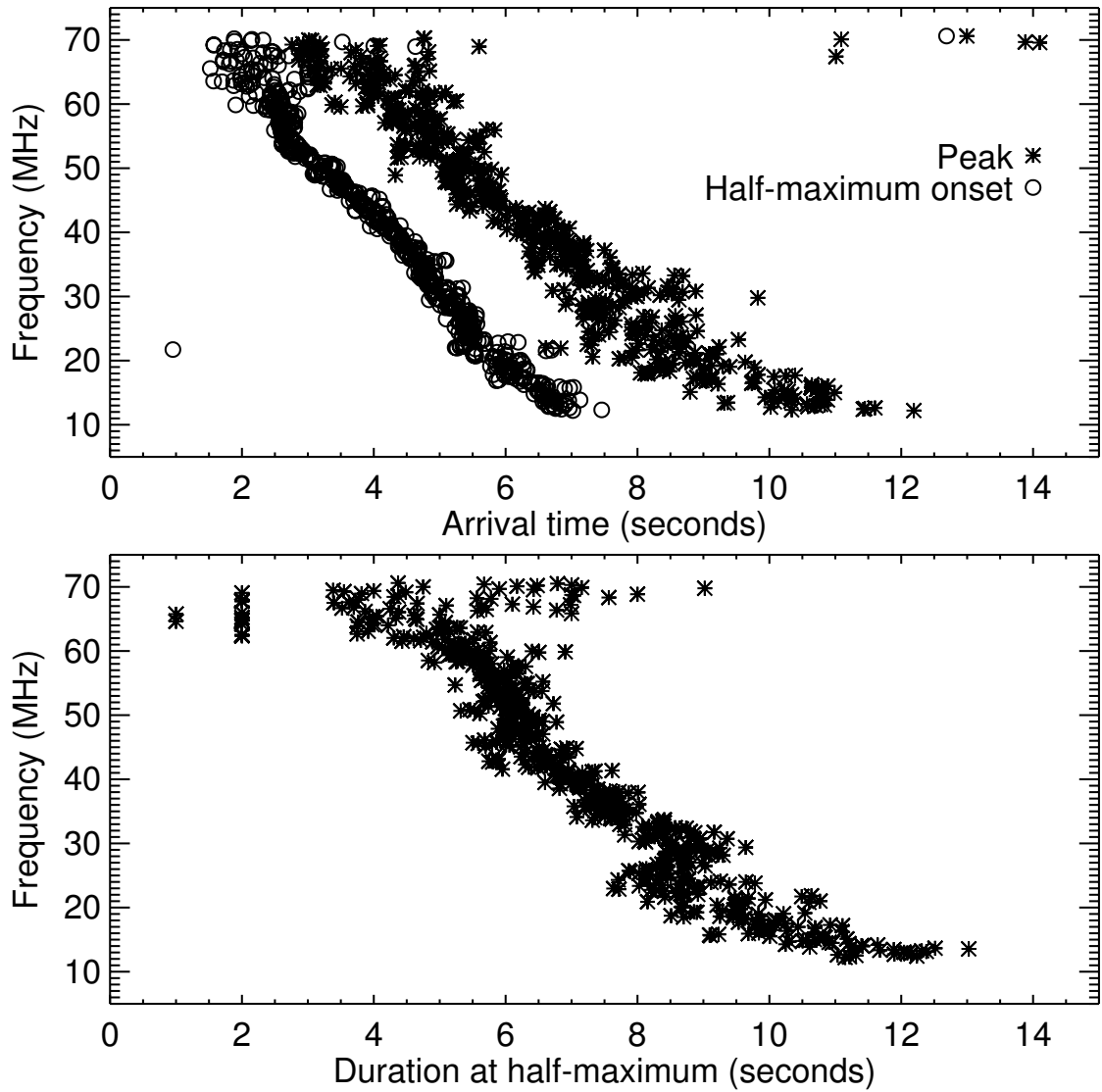

*Figure 2.* This figure plots fits of the temporal properties of the Type III burst in Fig. 1 as a function of frequency. The upper panel shows the time delay (after 16:19:00) of the arrival of Type III emission as a function of frequency, with circle symbols representing the time at which the burst reaches 50% of its maximum level at each frequency in the rise phase, and the asterisks showing the time of the peak emission. The lower panel shows the duration of the burst, measured between the half–maximum points at each frequency.

An example of the quality of the data is shown in Figure 1, which is an isolated Type III burst observed on December 7, 2006. The display range has been chosen so that the background can be seen in addition to the burst: in processing of the spectrum, narrow line features have been removed as well as a smooth background (containing the galactic background at low frequencies as well as some residual broader interference features). The final spectrum shows no prominent interference features despite covering a low–frequency range generally notorious for interference. The most prominent artefacts are residual varying emission near 20 MHz from an interference source, and the fall–off in the response of the dipole feed at high frequencies. The most common problem we find at the Green Bank site is a sporadic local broadband emission strongly modulated in frequency, but it was not present during this event. In summer, sporadic ionospheric E–layers are also commonplace in West Virginia and we see their effects in the presence of a digital TV signal in the spectra from 60–65 MHz, believed to be reflected in from Pittsburgh some distance away.

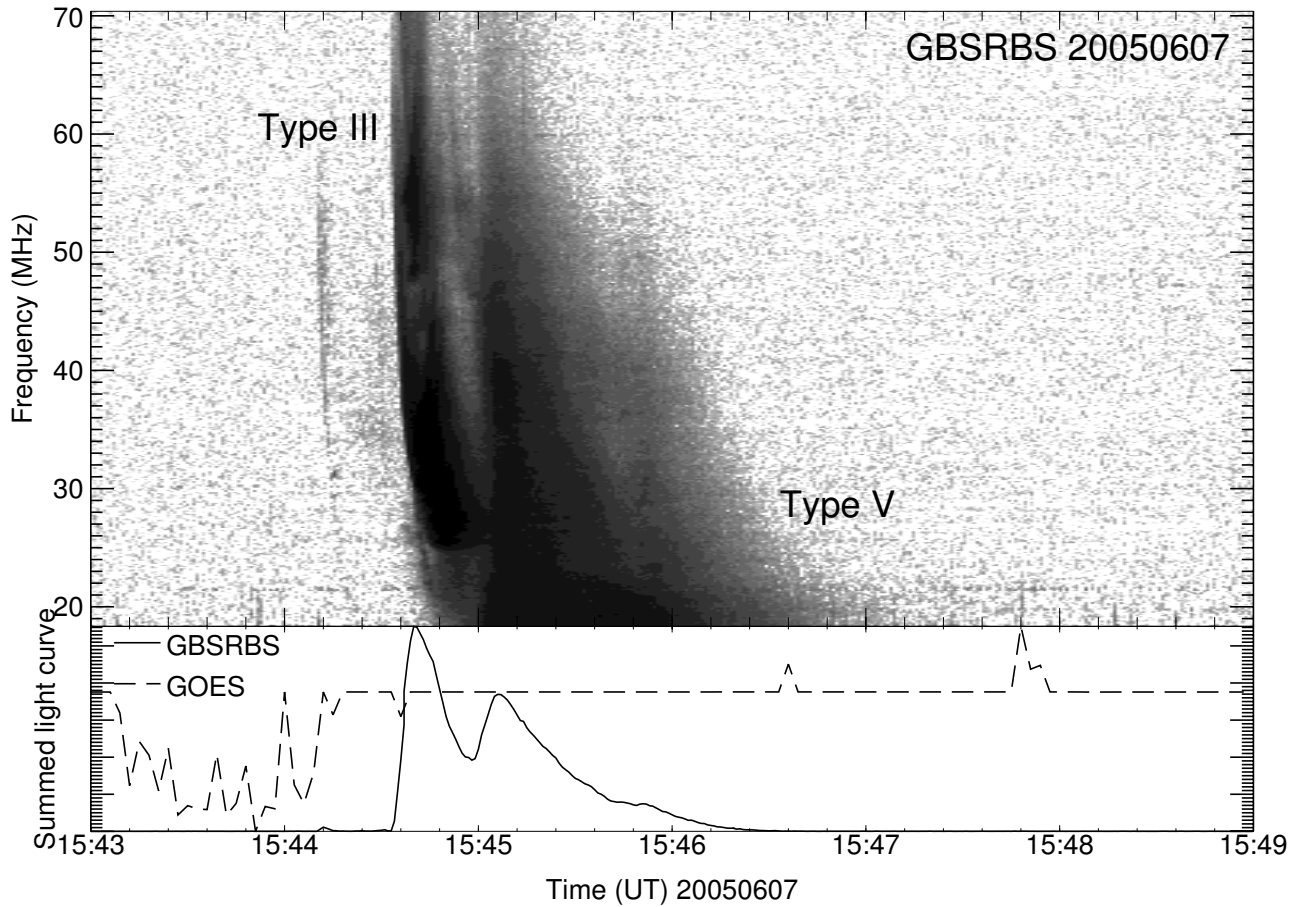

*Figure 3.* An example of a Type V burst, typically described as the extended phase of a Type III burst. In this example a Type III is observed as a precursor at about 15:44:30, and the Type V starts at 15:45:00 and lasts until 15:46:30 at 20 MHz.

## 4. Solar Radio Bursts

Solar radio bursts were amongst the first phenomena identified as targets for radio astronomy. Solar radio bursts at frequencies below a few hundred MHz were classified into 5 types in the 1960s (Wild et al., 1963). For Space Weather studies, three of the burst types are most relevant, and we discuss each of these separately below: Types II, III, and IV. Here we mention that Type I bursts are a non–flare–related phenomenon, consisting of a continuum component and a burst component. The continuum, also often referred to as a "noise storm", typically covers the frequency range 100–400 MHz, with variations on timescales of hours. The long duration of this emission suggests that it is due to energetic electrons trapped on closed coronal magnetic field lines. The associated Type I bursts are brief (of order seconds duration) and very narrow band, and tend to occur in drifting chains of 10–20 MHz bandwidth. Because they are not clearly associated with energy releases visible in other wavelength ranges, noise storms are an intriguing sign that energy release can continue in the corona on long timescales, but the absence of diagnostics at other wavelengths also makes them very difficult to study. Imaging observations with the Nançay Radio Heliograph generally show compact radio sources in the appropriate frequency range over most flare–productive active regions, even though they are not visible in radio dynamic spectra (which are not optimal for low–level broadband features), indicating that such sources, and the necessary associated energy releases, are quite common in the active solar corona.

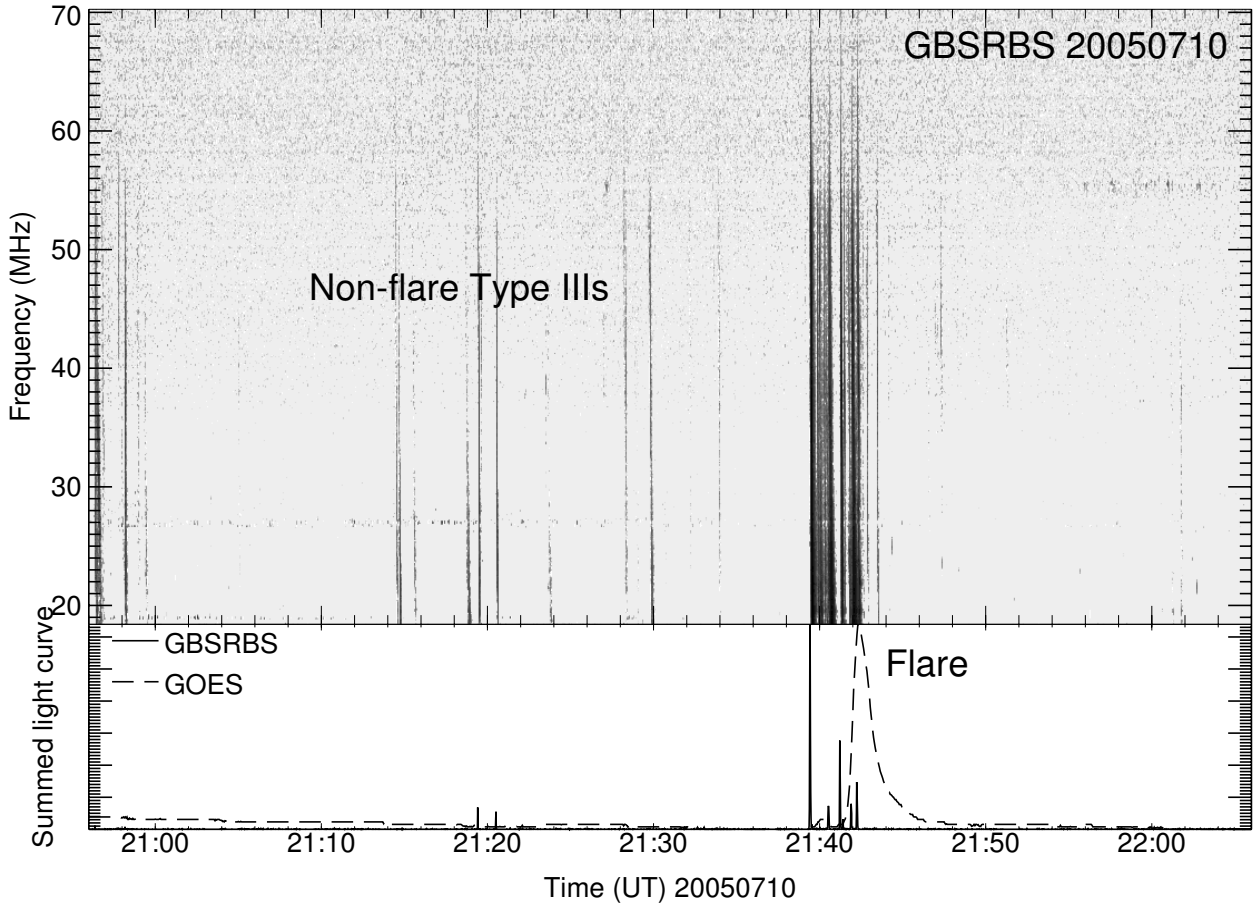

*Figure 4.* Dynamic spectrum of over an hour on a day of activity, showing Type III events occurring both in conjunction with a solar flare (at 21:40 UT) and in the absence of flare activity (between 21:10 and 21:35 UT).

## 5. Type III bursts

Type III bursts are brief radio bursts that drift very rapidly in frequency versus time (Fig. 1). Because the emission is at the plasma frequency (or its harmonic), the drift in frequency with time can be directly converted into a drift from high to low ambient coronal density with time. Coronal density models can then be used to infer a velocity for the exciter. In Figure 2 we show measurements of the arrival time and duration of the burst in Fig. 1 as a function of frequency: it drifts from 50 to 20 MHz in about 3 seconds, or 10 MHz s$^{-1}$. (Drift rates at higher frequencies are faster, because the density scale heights lower in the atmosphere are smaller, and a disturbance at constant speed will therefore have a higher frequncy drift rate.) To take an example, in the Newkirk coronal density model (Newkirk, 1961) these two frequencies correspond to heights of 0.5 and 1.1 $R_\odot$, implying a speed of almost half the speed of light. The inferred speed clearly depends on which of the many coronal density models one applies, but the general result is that for (relatively) "fast–drift" bursts (Type III bursts have consistently the fastest drift rates of bursts at metric wavelengths), the exciter speeds tend to be of order one–tenth the speed of light, and accordingly the only plausible drivers for Type III bursts are beams of electrons of energies up to tens of keV. Such beams of electrons have long been known to be very efficient producers of electrostatic Langmuir waves via the bump–in–tail instability. They can be seen to start at densities corresponding to the very low corona (frequencies up to several GHz) and propagate through the SRBS frequency range all the way out to 1 AU, where their electrons can be detected in–situ by spacecraft in the solar wind.

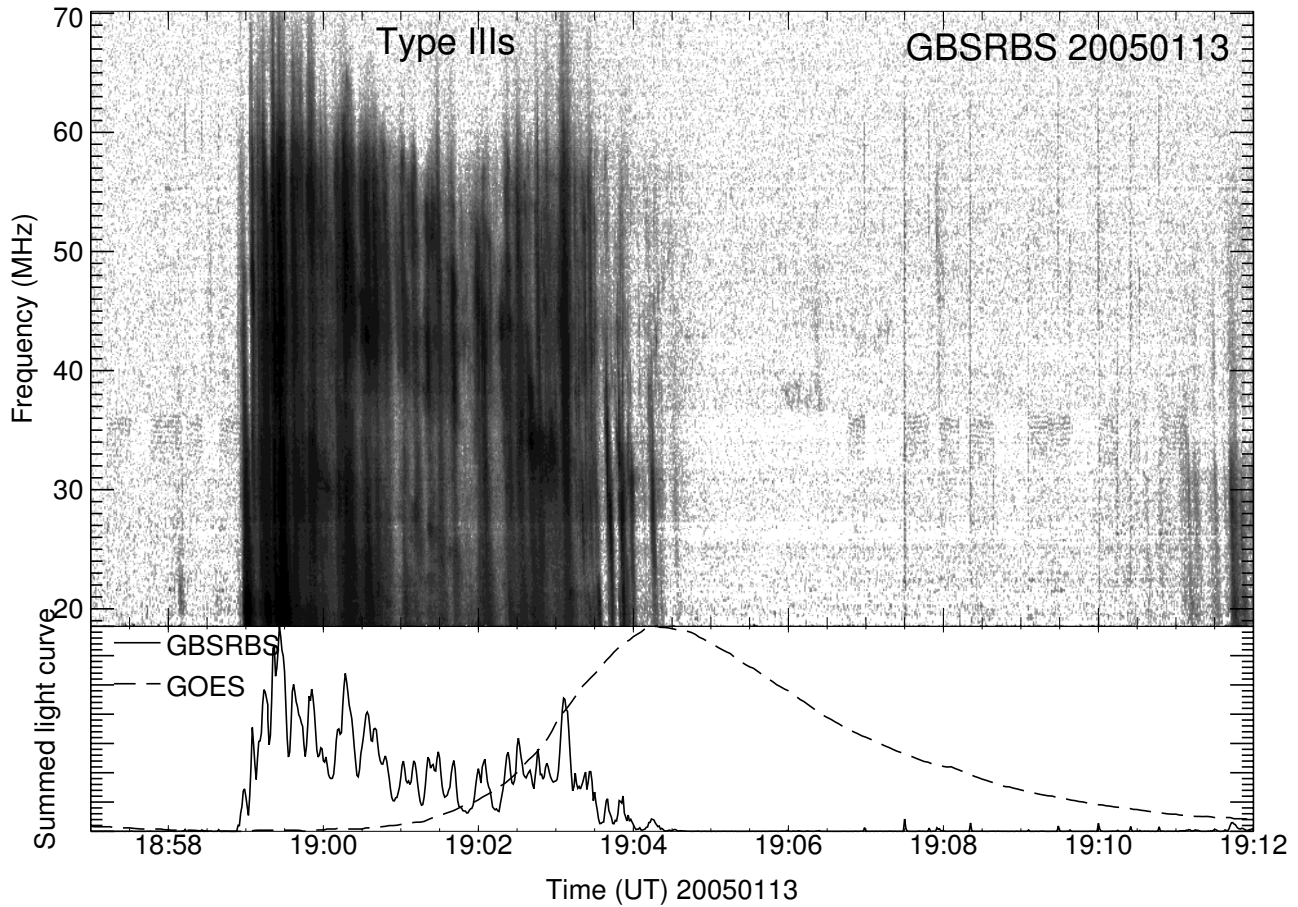

*Figure 5.* A large group of Type III bursts in the rise phase of a solar flare. The Type IIIs cease almost exactly at the time of the soft X–ray peak, regarded as the time at which energy release in the flare ends.

Associated with Type III bursts are the last of the original classes of meter–wavelength radio bursts, Type V. Figure 3 shows an example: the Type V emission is the longer–duration low–frequency component that looks to be "attached" to (and possibly emerging from) the decay phase of the Type III burst ahead of it. This is the defining characteristic of Type Vs: an extended phase following Type III emission at low frequencies, lasting for up to a minute. Type V bursts are often difficult to identify, particularly if there are other bursts present at the same time. They are relatively rare and have no known Space Weather implications.

In addition to isolated bursts such as Fig. 1, Type IIIs are commonly seen in the impulsive phase of solar flares, and the connection they imply between the acceleration region in solar flares and open field lines that reach the solar wind makes them important for understanding field line connectivity in flares and the access of flare–accelerated particles to the Earth. Figure 4 shows over an hour of data from an active period in which a large number of Type IIIs are seen. This figure demonstrates that large flares are not necessary to produce electron beams: Type III bursts can be seen at times when there is no activity at other wavelengths. The plasma emission mechanism is very efficient at converting a small amount of free energy in an electron velocity distribution into electromagnetic emission, so it does not require very many electrons in a beam to produce a detectable Type III burst.

Notwithstanding this fact, it is well established that a large fraction of flares, particularly impulsive flares (Cane and Reames, 1988a), exhibit Type III bursts at the onset. An example is shown in Figure 5, where Type III bursts occur repeatedly for almost the entire rise phase of the soft X–ray emission of a flare. The implication is that the energy releases that are responsible for the heating of the corona to soft X–ray temperatures ($10^7$ K) also result in the acceleration of electrons that escape as beams on open magnetic field lines, and so such Type III bursts potentially form a diagnostic of energy release and acceleration. Aschwanden et al. (1993) found cases where both upgoing and downgoing electron beams produced radio signatures, originating in a frequency range corresponding to a density of over $10^9$ cm$^{-3}$, and also suggested that the downgoing beams were well–correlated with particle precipitation into the solar chromosphere revealed by structure in the hard X–ray emission time profile, further connecting the Type III bursts to the main energy release events in flares.

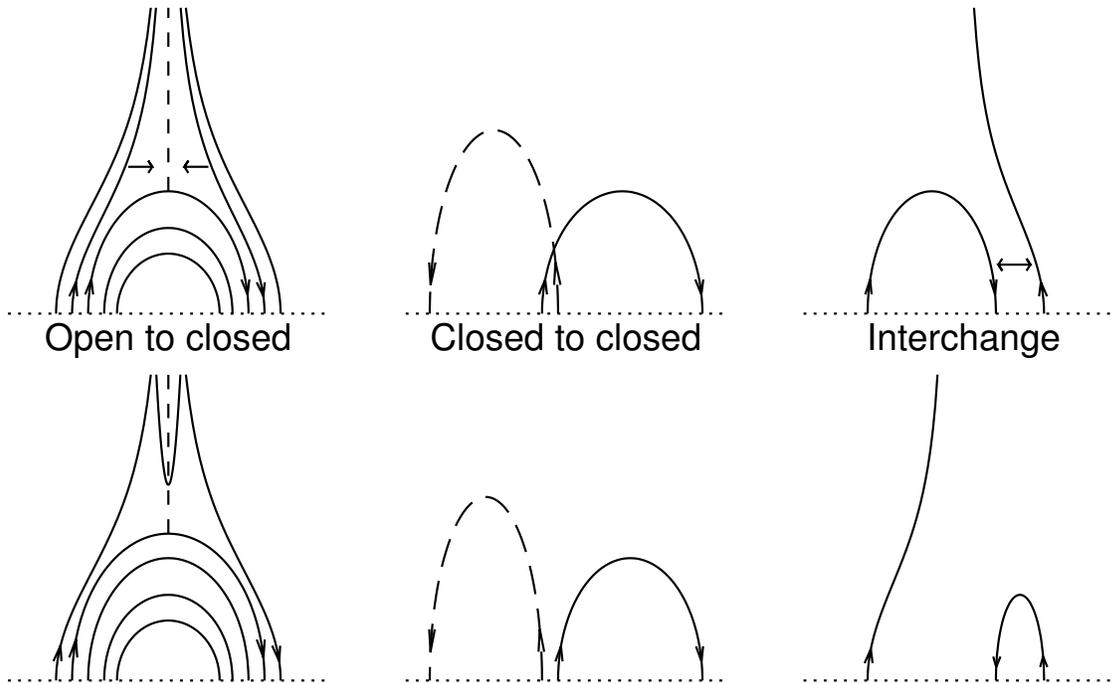

*Figure 6.* Magnetic reconnection topologies that may apply in flares. Three models are shown before (upper row) and after (lower row) reconnection. The left column shows reconnection of two open field lines in a "helmet–streamer" configuration to produce one closed field line and one open field line no longer connected to the solar surface at either end; the middle column shows the "quadrupolar" configuration in which two closed field lines reconnect to produce two new closed field lines; and the right column shows the case where one open and one closed field line swap topology.

The presence of Type III bursts generated by the energy release in flares has implications for flare models, and therefore for Space Weather prediction, because the electron beams are seen to propagate out to regions of very low density and therefore must be on open field lines. We believe that flare energy release involves magnetic energy density stored in the corona in a form, such as sheared non–potential magnetic fields, that is available for conversion to particle energy (thermal and nonthermal) because no other medium that we know of offers the possibility of storing sufficient energy to explain large solar flares (Wu et al., 1989). Conversion of magnetic energy into particle energy may take place as a by–product of the mechanism of magnetic reconnection, in which magnetic field lines may change topology and connectivity (field lines may be cut and then reconnected to other field lines, a non–ideal MHD process). The strongest magnetic fields in the solar photosphere occur in sunspots: most of the field lines passing through strong field regions appear to close within the solar atmosphere (i.e., both footpoints are connected to the solar photosphere), which suggests that the energy available for flares is probably stored on closed field lines. However, that need not be the case and open field lines (which have one end rooted in the solar photosphere and the other pulled out into the solar wind) also may be involved. A popular model for long–duration flares and the associated coronal mass ejections involves a so–called "helmet–streamer" configuration (Figure 6) in which closed loop field lines lie under a current sheet which separates regions of open field lines of opposite magnetic polarity: reconnection at the top of the loop creates new closed field lines (as well as field lines not rooted in the solar surface) and releases energy (see the discussion of flare models in Aschwanden, 2004). In this model open field lines are an intrinsic part of the energy release process, and it seems natural that particles accelerated in the energy release would find their way onto open field lines. However, Type III bursts are mostly found in the flash phase of very impulsive flares (Cane and Reames, 1988a), which are not thought to be due to the helmet–streamer model. Other models applied to very impulsive flares involve interactions between pairs of closed loops (the middle column of Fig. 6), such as when loops carrying magnetic flux emerging from beneath the solar photosphere reconnect with pre–existing loops in the solar corona. In such models the outcome of reconnection is two new closed loops, and then no open field lines are involved, making it difficult to see how Type III bursts could originate in the energy release region. Another class of flare models involves reconnection between one open field line and one closed field line, in which topology is switched, and in this model we again have open field lines as an intrinsic part of energy release.

In addition, a class of Type III bursts that occurs somewhat after (typically 10 minutes or so, during Type II emission) the impulsive phase has been reported to show a very high degree of association with solar energetic particle events (Cane et al., 2002). These bursts occur in groups and are labelled Type III-*l*; they are commonly seen below 14 MHz as well as in the SRBS range. Their association with solar energetic particle (SEP) events has made them an important topic of current studies. However, we have not seen any clear examples of this type of burst with SRBS so we do not present an example here. Initially it was thought that these fast drift bursts might originate in a Type II shock, since they often occur at the same time (i.e., well after the onset of the impulsive phase, hence a phenomenon quite different from the impulsive–phase Type IIIs discussed above) and in a similar frequency range (Dulk et al., 2000).

Cane et al. (2002) found that nearly all solar energetic proton events are preceded by such groups of Type III-*l* bursts, and that they are particularly prominent in dynamic spectra below the ionospheric cutoff at about 10 MHz. Since the Type III bursts in a Type III-*l* group often started at frequencies above the Type II emission visible in dynamic spectra at the same time, they argued that the Type III-*l*s had to originate lower in the corona than the Type II shock, and thus they were unlikely to originate in the Type II shock or in any shock associated with a fast CME, which is likely to be ahead of the Type II emission. This implies that the source of the Type III-*l* emission and, by implication, any associated energetic protons, was more likely to be in the flare region, and that open field lines must connect the acceleration region to the solar wind. This picture is in contrast to the belief that large gradual solar energetic particle (SEP) events are due to acceleration of particles by large coronal mass ejections (Reames, 1999). The association of Type III-*l* with SEPs is still being investigated and the exact relationship is by no means settled, but if it holds up then their observation, occurring up to tens of minutes before the arrival of the corresponding SEPs at the Earth, would hold promise as a Space Weather diagnostic.

## 6. Type II bursts

Type II bursts typically occur at around the time of the soft X–ray peak in a solar flare and are identified by a slow drift to lower frequencies with time in dynamic spectra, the frequent presence of both fundamental and second–harmonic bands (with a frequency ratio of 2), and splitting of each of these bands into two traces. The frequency drift rate is typically two orders of magnitude slower than that of the ("fast–drift") Type III bursts, so the two burst types are readily distinguished. Figure 7 shows a classic example of a flare in which Type III bursts occur at the onset of the event and a Type II burst is seen to start near the peak of the soft X–ray emission. Both fundamental and harmonic lanes are visible, and both show splitting. In rare events, a third harmonic trace can also be seen (Zlotnik et al., 1998).

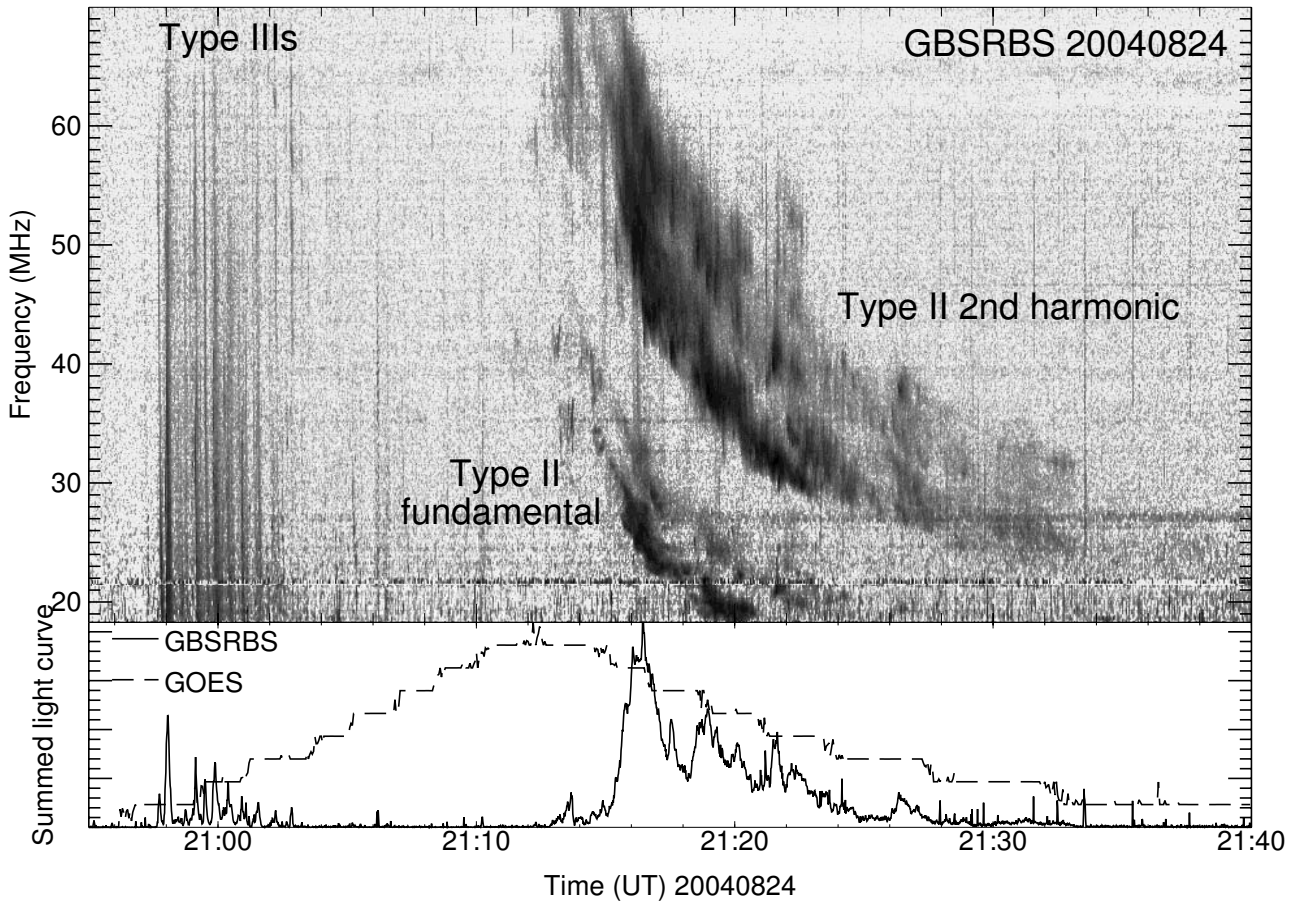

*Figure 7.* Classic Type III/II event: in the early rise of the impulsive phase of the flare (shown here by the onset of soft X–rays seen by the GOES satellite in the lower panel) a group of fast–drift Type III bursts is seen, followed after the soft X–ray peak by a fundamental–harmonic Type II burst with both lanes showing obvious splitting. This was a small (GOES class B5) flare.

The emission mechanism of Type II bursts is assumed to be plasma emission at the plasma frequency and its harmonic. The observed frequency drift rate can be converted into a velocity if the dependence of electron density $n_e$ on height is known, and it is found that a typical speed is of order 1000 km s$^{-1}$, or somewhat larger than the typical Alfvén speed in the corona. For this reason Type II bursts are agreed to be evidence for shocks in the corona, rendered visible by the radiation of electrons that they accelerate. There is almost always a delay between the flare onset and the start of Type II emission, which is attributed to the variation of the Alfvén speed with height in the corona: the Alfvén speed $v_A$ ($\propto B/\sqrt{n_e}$) is high in the low corona where magnetic field strengths are large, and decreases with height because $B$ typically falls off much more quickly with height than does density. A shock cannot form unless a disturbance exceeds the local magnetosonic fast mode speed, which is essentially $v_A$ in the low–plasma–$\beta$ solar corona. As height increases, $v_A$ decreases and the Mach number of a disturbance moving at a constant speed increases, producing a stronger shock.

The shocks that produce Type II emission have never been unambiguously identified at other wavelengths, although possible associations have been suggested and include coronal mass ejections, Moreton waves, and soft X–ray ejecta. Type II bursts are always seen in conjunction with flares, even though some of those flares are very small events, and there is a very healthy controversy as to whether the shocks are driven by CMEs or by some other flare phenomenon: with the improved coverage and sensitivity of coronagraphs in recent years, the correlation between Type II bursts and the presence of a coronal mass ejection (CME) has become increasingly tight, lending support to the idea that the shocks that produce Type II bursts are being driven by CMEs, without resolving the issue (Cliver et al., 1999; Gopalswamy et al., 1998; Cliver, 1999; Gopalswamy, 2000; Shanmugaraju et al., 2003; Cliver et al., 2004; Cane and Erickson, 2005). The particle acceleration exhibited by Type–II driving shocks, and their associations with flares and/or CMEs, make them important for Space Weather studies.

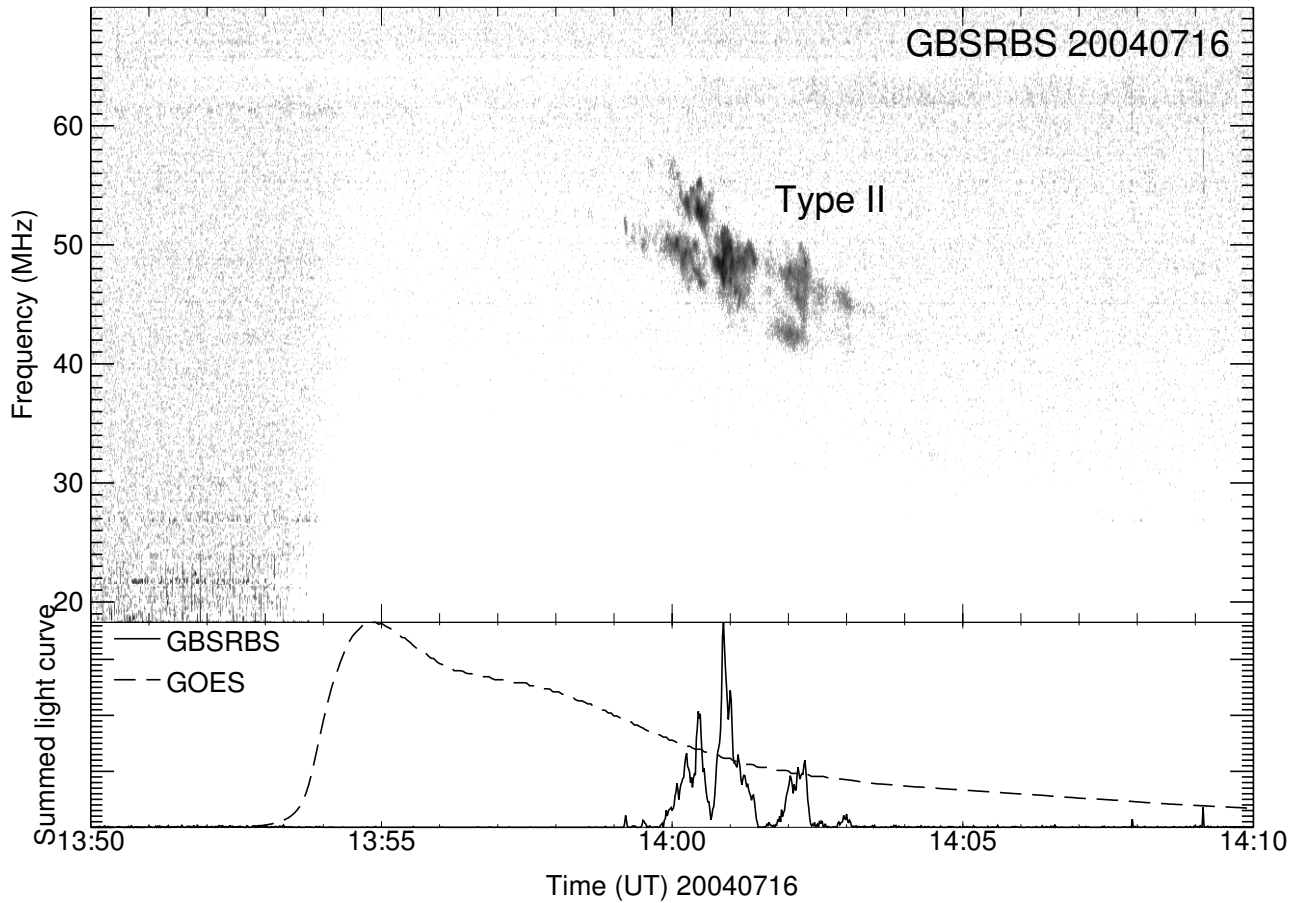

*Figure 8.* A ghostly Type II burst from a very large X3 flare. This flare produced a "short–wave fadeout": the large flux of extreme ultraviolet and soft X–ray photons early in the flare produced additional ionization in the ionosphere, which increased the absorption of radio emission from the Sun passing through the ionosphere. This absorption is visible in the figure in the lack of background noise starting near 13:53 UT at low frequencies, extending up to 60 MHz at the peak of the X–ray emission. A short section of one split lane of Type II emission is visible through this absorption from 13:59 to 14:03 UT.

Type II bursts are generally a low–frequency phenomenon, with most being detected below 100 MHz. In the first 18 months of operation of SRBS, only 1 Type II event was reported in SRBS's time range that was not seen below 70 MHz by SRBS. Of the 31 Type IIs observed by SRBS in this period, 3 events did not produce detectable CMEs (i.e., the images from the LASCO coronagraph on the SOHO satellite cover the interval of the event but do not show an eruption). The largest of the events is shown in Figure 8: this was a GOES class X4 flare which produced a short wave fadeout (due to additional absorption in the D layer of the ionosphere resulting from the ionizing EUV and soft X–ray flux of the flare). This fadeout may obscure some of the radio emission from this event, but a short period of Type II emission, with the normal frequency drift rate and showing split–band features, is clearly present.

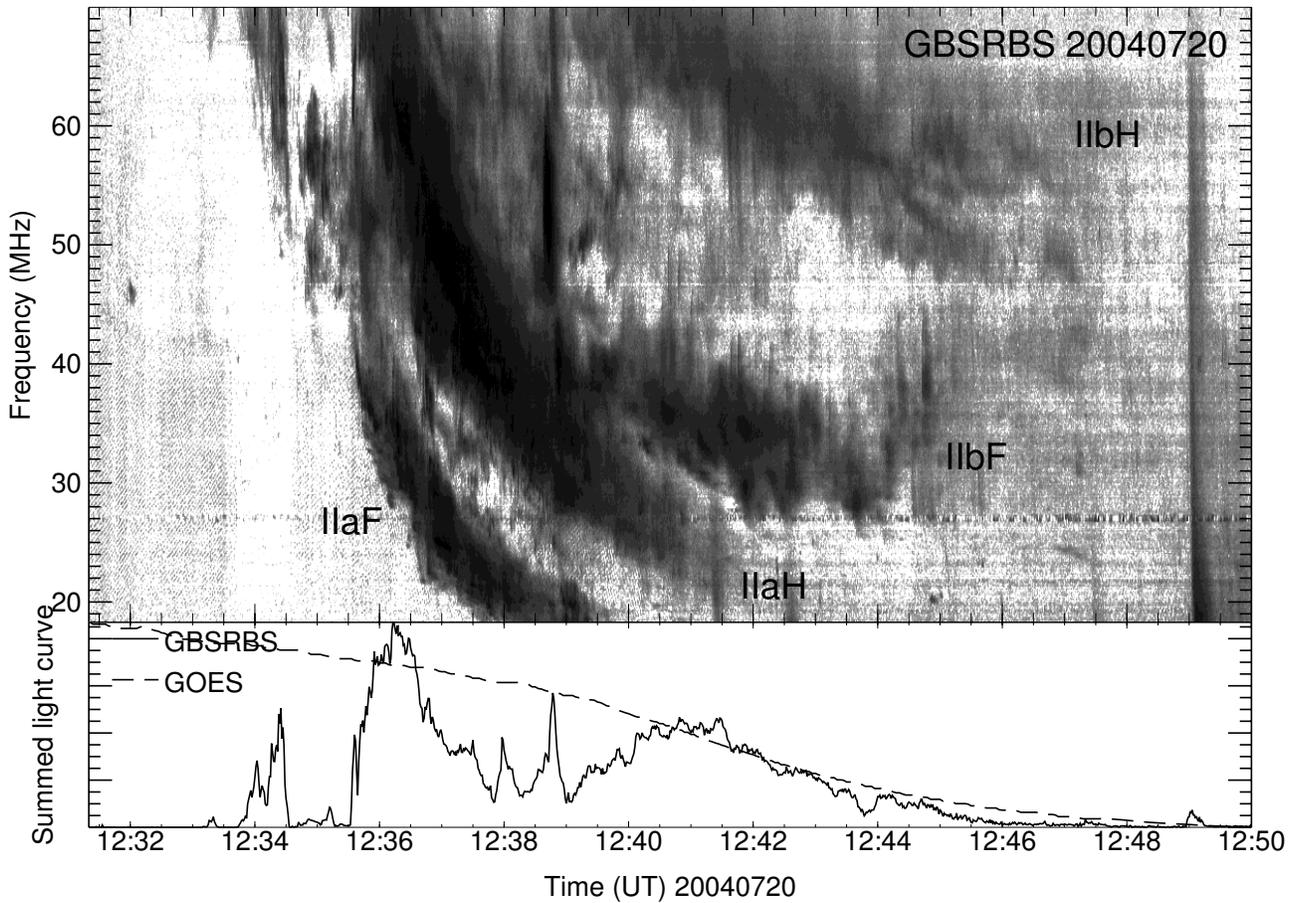

*Figure 9.* An event in which two separate traces of fundamental–harmonic Type II emission are visible simultaneously. The pairs are labelled IIa F (fundamental) and H (harmonic), IIb F and H. The two pairs show very different drift rates (IIa has the faster drift rate): an interpretation is that a large scale shock is propagating through the corona and producing Type II emission in two distinct regions with very different density gradients along the direction of propagation. This was a large (GOES class M9) solar flare with a coronal mass ejection.

Assessing the significance of the non–detection of a CME is complicated by viewing angle effects: because of the geometry of white–light scattering, CMEs occurring at the solar limb are much easier to detect in coronagraph images than are events that appear to be projected against the bright solar disk, and this effect will lower the degree of association of CMEs with Type IIs. Yashiro et al. (2005) conclude that half of all CMEs produced by GOES C–class flares were not detected by LASCO, but that detection of CMEs is complete for flares at GOES class X3 or larger, irrespective of the longitude of the event. This event occurred at a longitude of E35, and any CME should have been detectable by LASCO based on the study of Yashiro et al. (2005).

The standard picture of Type II bursts is of electrons accelerated by a broad shock front propagating through the corona: if the shock is being driven by a CME, then we know from coronagraph observations that the shock front should cover a large area. Given this picture, it seems odd that the Type II lanes are usually relatively narrowband and well–defined, since emission at many widely–separated locations in the corona should be at very different electron densities, and thus cover a wide frequency range at any given instant. Electron acceleration might not occur everywhere on the shock front: for example, it is quite possible that acceleration only takes place where the shock front is orthogonal to the local magnetic field, and this would limit the range of locations in which emission occurs. More complex events are seen: in particular, Type IIs in which two sets of harmonic lanes are seen simultaneously are not uncommon. An example is shown in Figure 9, in which the two pairs of lanes clearly have different drift rates. These could indicate emission from two different locations within a large–scale shock, such as we would expect from a monolithic disturbance such as a CME, or alternatively they could be due to the presence of two separate, more localized shocks in this event.

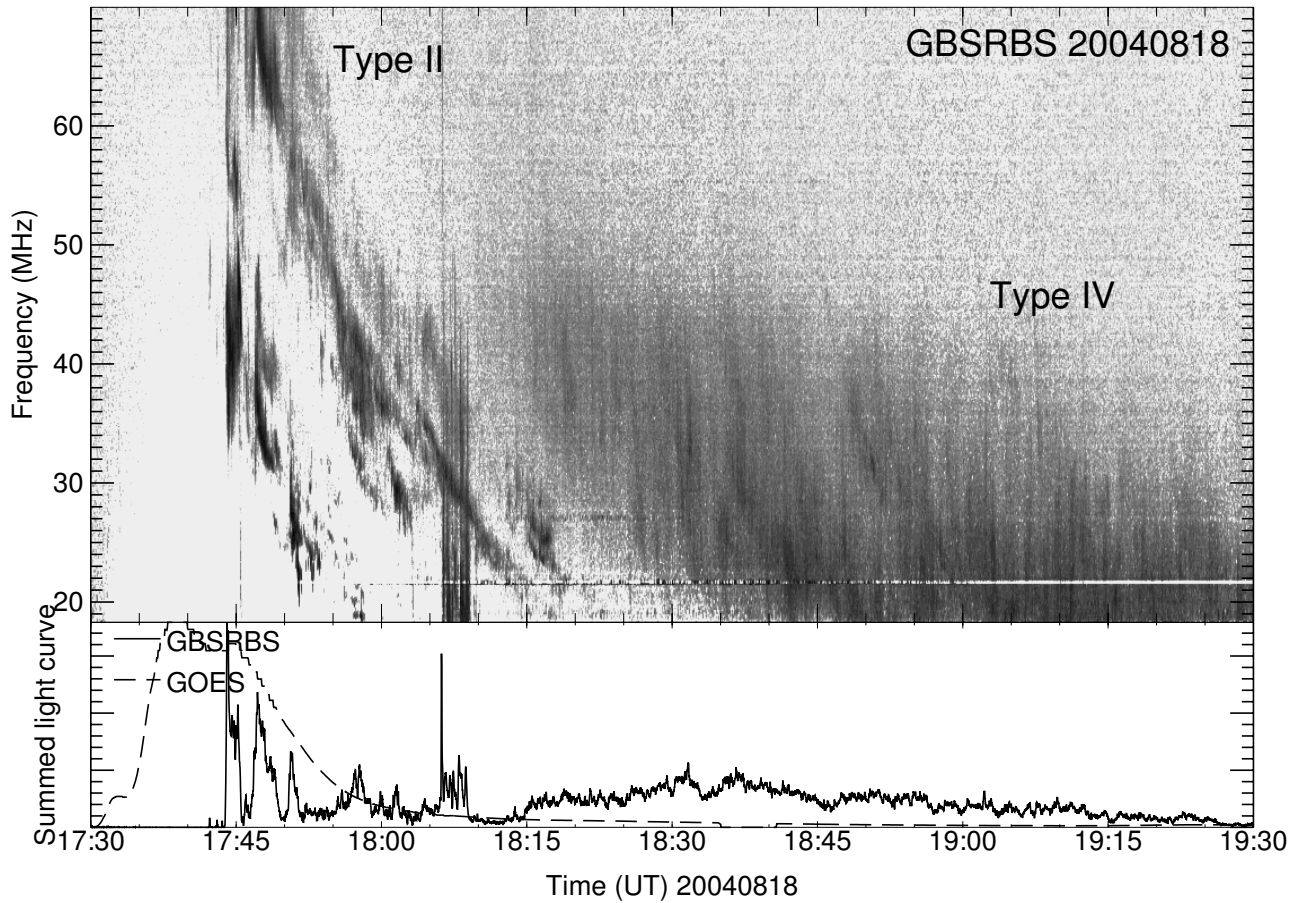

*Figure 10.* A classic Type II–IV event from a large (GOES class X2) flare. In this class of event there are often no Type III bursts in the impulsive phase: a Type II burst occurs at around the time of the soft X–ray maximum, and is then followed by a broadband Type IV emission which, as in this case, may drift to lower frequencies with time. Note that a short–wave fadeout is also evident in this event as a decrease in the apparent background noise level from 17:32 UT onwards.

There continues to be a great deal of research on the relationship between Type II radio bursts, CMEs and particle acceleration, and the Space Weather applications remain unclear.

## 7. Type IV bursts

Type IV bursts are broadband quasi–continuum features associated with the decay phase of solar flares. They are attributed to electrons trapped in closed field lines in the post–flare arcades produced by flares; their presence implies ongoing acceleration somewhere in these arcades, possibly at the tops of the loops in a "helmet–streamer" configuration. Although they are not as common as Type II and III bursts, and there has been somewhat less work on their properties (Cane and Reames, 1988a; LaBelle et al., 2003; Zlotnik et al., 2003; Magdalenić et al., 2005; Pick et al., 2005), Type IV bursts have long been of interest in Space Weather studies because they have a high degree of association with solar energetic particle events.

Figure 10 shows a stereotypical example of a Type IV burst: in this event there are no Type III bursts at onset (although a significant short–wave fadeout is present) and the first meter–wavelength radio emission seen is a Type II burst at the peak of the soft X–ray emission. As the Type II moves down towards the ionospheric cutoff, a broadband continuum appears with substructure, notably broadband modulations on timescales of seconds resembling fast–drift bursts. This is a Type IV burst, and as in this case it can last for an hour or longer. Unlike Type II events, Type IV events are almost always associated with large flares, and usually those with long durations (Cane and Reames, 1988a). Cane and Reames (1988b) also showed that Type IV events do tend to be associated with Type II events: in their study 88% of over 200 Type IV bursts observed by the Culgoora Observatory were preceded by Type II events, whereas only about one–third of all Type II bursts are followed by a Type IV burst. This event was associated with a CME at the west limb, with the relatively low speed of 600 km s$^{-1}$, but no energetic particle event was detected.

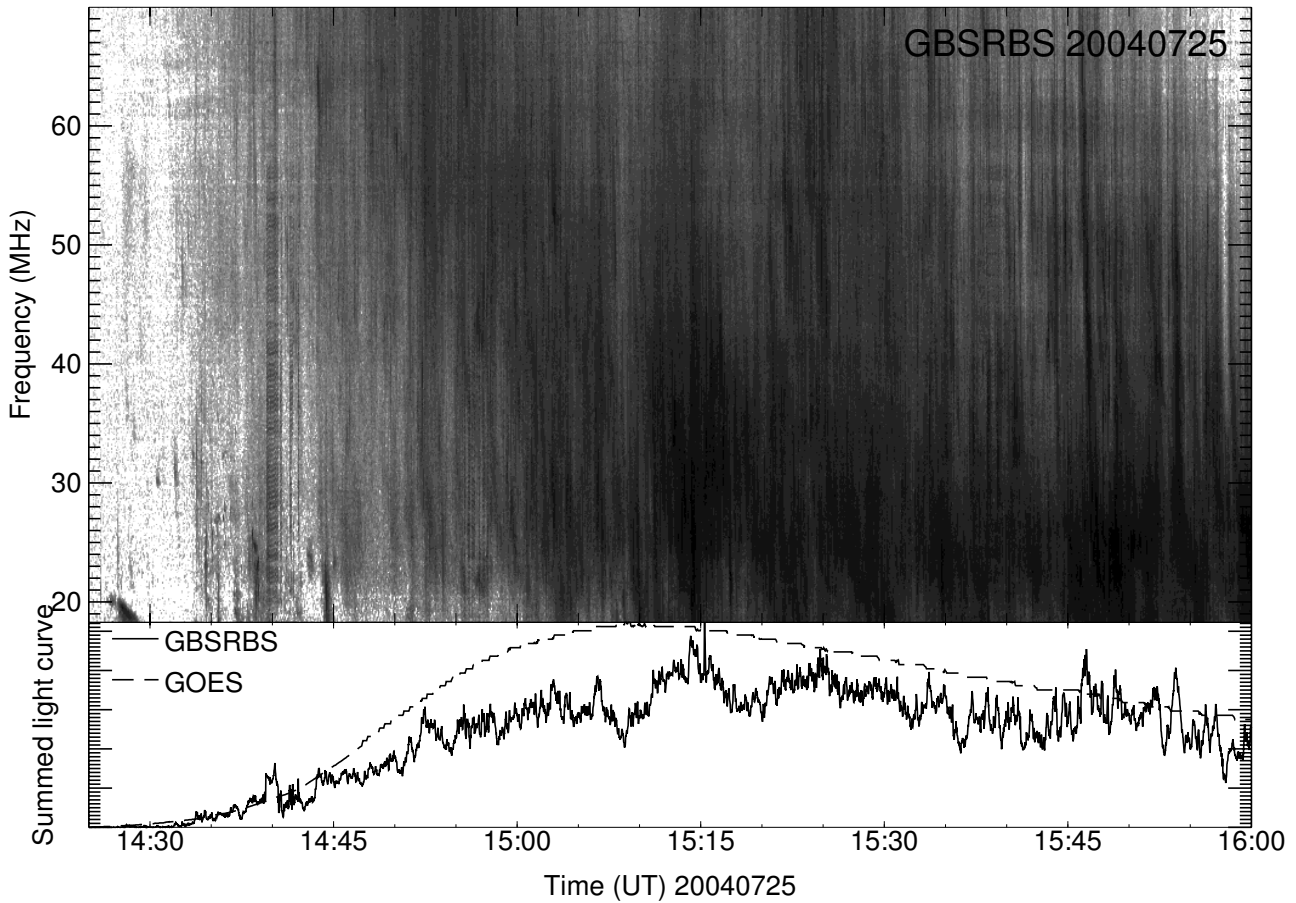

*Figure 11.* An example of a Type IV burst associated with a GOES class M1 flare. The burst is a broadband emission lasting for several hours, but with modulation in time that can resemble a large group of individual fast–drift bursts.

Despite the high correlation of Type IV bursts with Type II bursts, the production of Type IV bursts seems to be a peculiarity associated with certain active regions. A remarkable example occurred with AR 10786 in July, 2005: after July 5, this active region produced at least 6 recorded Type IV bursts and not one of them was preceded by a Type II burst. Regular observers notice that Type IV bursts tend to be associated with certain active regions: other regions may be prolific producers of Type II bursts and never produce Type IVs.

Figure 11 shows another example of a long–duration Type IV in which the rapid modulation and the presence of substructure are clearly visible. Substructure within Type IV bursts is, in fact, a major source of confusion for observers identifying Type II bursts: inspection of dynamic spectra suggests that in the first 18 months of operation of SRBS, 27 Type II bursts were reported in *Solar Geophysical*

*Data* and of these 3 were not Type II bursts, but rather were structures within the Type IV bursts which resembled slow–drift features, but occurring well after the impulsive phase of the associated flares. The properties of and models for the substructure in Type IV bursts have been discussed by Aurass et al. (2003) and Zlotnik et al. (2003). The broadband and non–drifting nature of Type IV emission has led to the widespread belief that they are due to electrons trapped on closed magnetic field lines. In such a geometry, short–duration modulations can be produced by repeated injections of energetic electrons into a loop. Such injections imply the occurrence of repeated small energy releases associated with the acceleration of such electrons. A natural location for the accelerator is at the top of the post–flare loop systems commonly associated with the long-duration flares that tend to produce Type IV bursts: the long gradual rise of such loop systems and their ongoing soft X–ray emission have led to the idea that they experience energy release that is distinct from the impulsive energy release responsible for the flare itself, and some portion of such an energy release could be used to accelerate electrons that then produce Type IV emission on closed field lines in the arcade. Imaging observations at metric wavelengths tend to show that most Type IV bursts are stationary in the corona, but the association with the top of post–flare loop arcades could not be confirmed directly due to the low spatial resolution typical of long wavelengths. However, in the well–known event of 2002 April 21 Kundu et al. (2004) were able to show that there was non–thermal gyrosynchrotron emission at microwave frequencies (where high–spatial–resolution images were available) at the base of the post–flare arcade in that event at the same time as Type IV emission was seen in a dynamic spectrum at decimetric wavelengths, consistent with the general picture of Type IV events.

A sub–class of Type IV bursts, known as "moving Type IVs", is of great interest because they are seen to move in images at speeds similar to CME speeds, but they are difficult to separate from stationary Type IV events just from their spectral characteristics alone. Moving Type IV bursts are predominantly a low–frequency phenomenon, occurring at frequencies below 100 MHz, and due to the scarcity of imaging capability at such low frequencies presently (soon to be remedied by the construction of the Long Wavelength Array in the U.S.), they have been almost impossible to identify for some time. For this reason, and because of their relative rarity, there has been little research on them in recent years.

## 8.  Other bursts

Figure 12 shows another class of burst less often discussed than the burst types above and much harder to identify. This is a continuum emission occurring during the impulsive phase of a flare. In this event it takes the form of an extremely bright low–frequency broadband continuum, showing some modulation in time but generally little characteristic structure in the dynamic spectrum. It was produced by a large (GOES class M8) flare. Originally all broadband continua were called Type IV bursts, but over the years it was realized that this was inadequate for the range of characteristics such continua exhibit, and so many subclasses of Type IV burst were developed that the term was no longer useful (see the comprehensive historical discussion by Robinson, 1985). The term "Type IV" burst is now generally reserved for the two classes of event described in the preceding section: continua occurring after the impulsive phase of a flare, generally after a Type II event and stationary in the corona, and moving Type IV events which appear similar in dynamic spectra but move rapidly outwards. Other emissions occurring in the impulsive phase (i.e., during the rise and peak of the soft X–ray emission from the event) that are not clearly fast drift Type III bursts or slower–drift well defined Type II bursts now tend to be called "flare continuum", or "flare continuum early" in the terminology of Robinson (1985), who wanted to distinguish it from broadband continua seen later in flares. In practice, there seems to be little in the way of characteristic features that identify flare continuum apart from its timing relative to the impulsive phase and its broadband nature, and (speaking personally) it seems to be more useful as a designation of what an emission is not (i.e., the other well known burst types), rather than what it is.

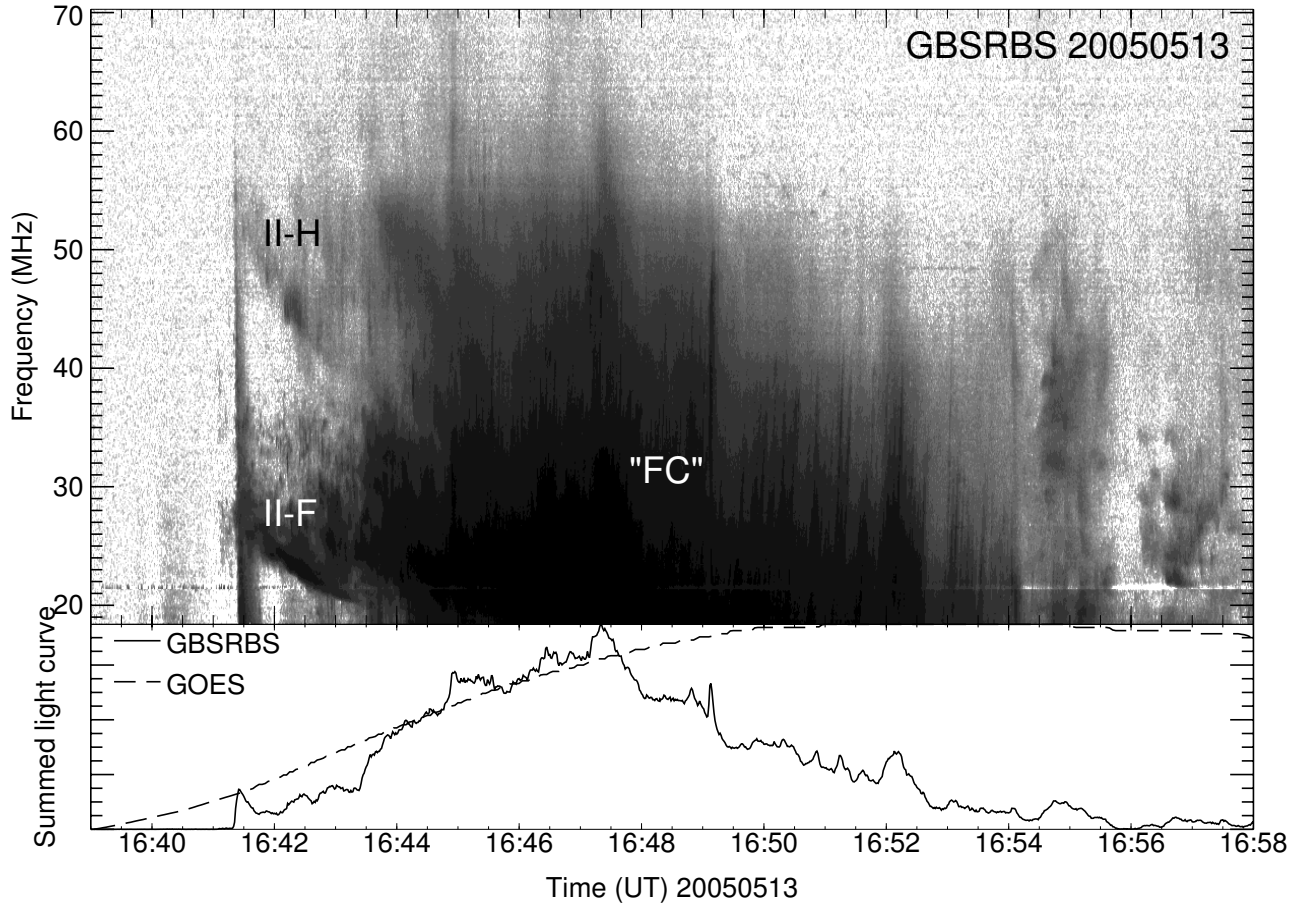

*Figure 12.* An extremely bright (but difficult to classify) radio burst. This was from a GOES class X1 flare, and was followed several minutes later by a Type IV burst clearly separated from the emission shown here. There may be a fundamental–harmonic pair feature with drift rate similar to a Type II burst in the earliest stage, at 16:42 UT (labelled II-F and II-H). The subsequent emission (labelled "FC") gets brighter as frequency decreases. This may be a form of "early flare continuum" (Robinson 1985).

The event shown in Fig. 12 occurred near disk center in conjunction with a very fast halo CME and a gradual–rise solar proton event. It was also associated with a major interplanetary radio burst (consistent with the bright low–frequency emission in Fig. 12) and emission seen all the way out to the Earth's orbit. Thus, this was an event with significant Space Weather effects, but at present there is no particular reason to view the flare continuum emission in the dynamic spectrum of this one event as a significant diagnostic.

One other Space Weather radio diagnostic that may become more important is the detection of synchrotron emission from the leading edge of coronal mass ejections. This possibility was demonstrated by Bastian et al. (2001) using images from the Nançay Radio Heliograph: at frequencies from 169 to 408 MHz, they found a beautiful arc–shaped feature in radio images, expanding as it moved outwards at a speed matching that of a CME visible in coronagraph data, and coincident with it in the single coronagraph image taken while the radio source was visible. From the radio spectrum and the relatively weak brightness temperature of the emission, Bastian et al. (2001) inferred that the emission mechanism was synchrotron emission in the magnetic fields associated with the coronal mass ejection from electrons accelerated by a shock driven into the solar wind by the CME. The emission was not visible in dynamic spectra of the radio event, and indeed, due to the limited dynamic range of the radio images, probably would not have been visible if the normally much brighter plasma emission sources low in the corona were not suppressed because the flare region was beyond the limb. Such data offer the possibility of directly imaging the locations of acceleration by shocks driven by CMEs, and further data are required to determine the relationship between the accelerated electrons seen in the radio images and the energetic protons that pose a threat to astronauts and to satellite operations, but as more low–frequency imaging facilities become available, with the possibility of imaging at frequencies selected to be devoid of bright plasma emission sources, this promises to be a productive area of Space Weather research.

## 9. Conclusions

The nature of low–frequency solar radio bursts and their potential for the study of Space Weather has been discussed. We have shown that Type III bursts are indicators of acceleration of electrons, and of the access of those electrons to open field lines, i.e., magnetic field lines in the corona that do not close within the solar atmosphere but instead become part of the solar wind and potentially feed energetic particles to the Earth's location. The common occurrence of Type III bursts early in the rise of impulsive solar flares may indicate that open field lines are an essential part of models for energy release by magnetic fields in such flares. Type II bursts are indicators of shocks in the solar atmosphere, and the nature of the driver of those shocks remains a subject of intense interest. There do seem to be some large events in which coronal mass ejections are not seen, when we would expect them to be easily detectable if present. CMEs clearly do produce detectable synchrotron emission that outlines the leading edge of the eruption, and this promises to be a valuable diagnostic in future imaging data. Type IV bursts continue to have a strong association with large long–duration flares that are often the source of energetic particle events in the solar wind, and they are evidence that particle acceleration continues in the solar corona long after the impulsive phase of a flare.

Solar radio observations will continue to play an important role in Space Weather studies because they are sensitive to the regions of the solar atmosphere in which many Space Weather phenomena originate. They can also see features that are not visible at other wavelengths, and thus complement other facilities such as satellite–borne coronagraph instrumentation.

## Acknowledgements

SRBS is supported by an NSF Division of Atmospheric Sciences grant under the National Space Weather program. Tim Bastian is the principal investigator for SRBS and Rich Bradley (both of NRAO) is the chief engineer. Scientific activities with SRBS are supported by NASA Living With a Star grant NNX06AC18G.